\documentclass[12pt,aps,article]{revtex4}

\usepackage{amsmath}
\usepackage{mathrsfs}
\usepackage{bm}
\usepackage[pdftex]{hyperref}

\begin{document}
\begin{titlepage}	
\title{Mathieu functions computational toolbox implemented in Matlab}
\author{E. Cojocaru}
\affiliation{Department of Theoretical Physics, 
Horia Hulubei National Institute of Physics and Nuclear Engineering, Magurele-Bucharest P.O.Box MG-6, 077125 Romania}

\email{ecojocaru@theory.nipne.ro}

\begin{abstract}
The Mathieu functions are used to solve analytically some problems in elliptical cylinder coordinates. A computational toolbox was implemented in Matlab. Since the notation and normalization for Mathieu functions vary in the literature, we have included sufficient material to make this presentation self contained. Thus, all formulas required to get the Mathieu functions are given explicitly. Following the outlines in this presentation, the Mathieu functions could be readily implemented in other computer programs and used in different domains. Tables of numerical values are provided. 
\end{abstract}

\maketitle
\end{titlepage}
\section{Introduction}
Some problems regarding the elliptical cylinders can be solved by using an analytical approach like that applied to circular cylinders: one separates the variables and the exact solution is given by expansions involving angular and radial Mathieu functions. These functions have been introduced by Emile Mathieu in 1868 by investigating the vibrating modes in an elliptic membrane \cite{1}. Details (tables or relations) concerning the Mathieu functions can be found for example in \cite{2,3,4,5,6,7,8,9,10,11,12}. For circular cylinders the solutions involve readily available trigonometric and Bessel functions, while for elliptical cylinders there are still controversial and incomplete algorithms for computing the Mathieu functions. Largely applied computer programs provide only few or none routines refering to the Mathieu functions. 
 
A computational toolbox for Mathieu functions was implemented in Matlab \cite{13}. Since not all people are familiarized with the Matlab program, in this presentation the mathematics is outlined. Tables of numerical values are provided. Following the outlines in this presentation, it would be a readily task to implement the Mathieu functions in other computer programs and use them in different domains.

One reason for the lack of algorithms for Mathieu functions was probably the complicated and various notation existent in the literature. A main purpose for us was to simplify as much as possible the notation. With a simplified and self-contained notation, the use of Mathieu functions should be as simple as the use of Bessel functions. We largely followed the notations used by Stratton \cite{6} and Stamnes \cite{11,12}, but we introduced further simplifications. Since the notation and normalization for Mathieu functions vary in the literature, we have included sufficient material to make this presentation self contained. Thus, all formulas required to get the Mathieu functions are given explicitly. Tables of numerical values are provided. Examples of Mathieu functions applied to plane wave scattering by elliptical cylinders are given in \cite{14,15}.  

\section{Fundamentals}

\subsection{Elliptical cylinder coordinates}

Let consider an ellipse in the plane $(x,y)$ defined by equation $(x/x_0)^2+(y/y_0)^2=1$ with $x_0>y_0$. The semifocal distance $f$ is given by $f^2=x_0^2-y_0^2$ and the eccentricity is $e=f/x_0<1$. The elliptic cylindrical coordinates $(u,v,z)$ are defined by relations
\begin{eqnarray}
\label{eq:1}
x=f\cosh u \,cos v,\qquad  y=f\sinh u\,\sin v,\qquad z=z
\end{eqnarray}
with $0 \leq u < \infty$ and $0 \leq v \leq 2\pi$. In terms of $(\xi ,\eta, z)$, with $\xi=\cosh u$ and $\eta=\cos v$, the elliptic cylindrical coordinates are defined by relations
\begin{eqnarray}
\label{eq:2}
x=f\xi\,\eta,\qquad  y=f\sqrt{(\xi^2-1)(1-\eta^2)},\qquad z=z .
\end{eqnarray}  
The contours of constant $u$ are confocal ellipses (of semiaxes $x_0=f\xi$, $y_0=f\sqrt{\xi^2-1}$) and those of constant $v$ are confocal hyperbolas. The $z$ axis coincides with the cylinder axis.
The scale factors $h_j$, with $j=\xi,\eta, z$, are defined like as for any coordinate transformation \cite{6},    
\begin{eqnarray}
\label{eq:3}
h_\xi=f\frac{\sqrt{\xi^2-\eta^2}}{\sqrt{\xi^2-1}}, \qquad 
h_\eta=f\frac{\sqrt{\xi^2-\eta^2}}{\sqrt{1-\eta^2}},  \qquad
h_z=1 .
\end{eqnarray}

\subsection{Wave equation in elliptic cylindrical coordinates}

The scalar wave equation $(\nabla^2+k^2)U(\bf{r})=0$, where $\bf{r}$ is the position vector, $k$ is the wave number, $k=2\pi\sqrt{ \epsilon}/\lambda$, $\epsilon$ is the permittivity, and $\lambda$ is the wavelength in vacuum, when expressed in elliptic cylindrical coordinates becomes
\begin{equation}
\label{eq:4}
\Big[\frac{2}{f^2 \left(\cosh 2u- \cos 2v \right)} \Big( \frac{\partial^2}{\partial {u}^2} + \frac{\partial^2}{\partial {v}^2} \Big) + \frac{\partial^2 }{\partial {z}^2}+k^2\Big] 
U(u,v,z) = 0 .
\end{equation}
Using a solution of the form $U=Z(z)S(v)R(u)$ gives 
\begin{eqnarray}
\label{eq:5}
\Big( \frac{\mathrm{d}^2}{\mathrm{d}z^2} 
 + k_z^2 \Big) Z(z) = 0   ,\\
\label{eq:6} 
\Big[ \frac{\mathrm{d}^2}{\mathrm{d}v^2} 
+ \left( a - 2 q \cos 2v \right) \Big] S(v) = 0 ,\\
\label{eq:7}
\Big[ \frac{\mathrm{d}^2}{\mathrm{d}u^2} 
- \left( a - 2 q \cosh 2u \right) \Big] R(u) = 0 ,
\end{eqnarray}
where $k_z$ is the wave vector component on $z$ direction, $q=k_{\tau}^2f^2/4$, with $k_{\tau}^2=k^2-k_z^2$, and $a$ is separation constant. Equation~(\ref{eq:5}) has solution $Z(z)=\exp{(ik_zz)}$. Equations~(\ref{eq:6}) and~(\ref{eq:7}) are known as the angular and radial Mathieu equations, respectively.

\section{Angular Mathieu functions}

In this presentation, only the periodic solutions of period $\pi$ or $2\pi$ are considered. For a given order $n$, there are four categories of periodic solutions satisfying (\ref{eq:6}),
\begin{eqnarray}
\label{eq:8}
1 \qquad \textrm{even-even:} \qquad  
S_{ee}(v,q,n)=\sum_{j=0}^\infty
A_{ee}^{(2j)}(q,n)\cos(2jv),\qquad \quad  \nonumber \\
2 \qquad \textrm{even-odd:} \qquad  
S_{eo}(v,q,n)=\sum_{j=0}^\infty
A_{eo}^{(2j+1)}(q,n)\cos[(2j+1)v], \\
3 \qquad \textrm{odd-even:} \qquad   
S_{oe}(v,q,n)=\sum_{j=1}^\infty
A_{oe}^{(2j)}(q,n)\sin(2jv),\qquad \quad \nonumber \\
4 \qquad \textrm{odd-odd:} \qquad  
S_{oo}(v,q,n)=\sum_{j=0}^\infty
A_{oo}^{(2j+1)}(q,n)\sin[(2j+1)v]. \nonumber  
\end{eqnarray}
$A_{pm}$ with $p,m=e,o$ are expansion coefficients. In the following, the angular Mathieu functions are denoted $S_{pm}(v,q,n)$, with $p,m=e,o$. Instead of two angular Mathieu functions, even $S_{ep}$ and odd $S_{op}$, with $p=e,o$ \cite{12}, a single angular Mathieu function $S_{pm}$, with $p,m=e,o$, is considered refering to all the four categories. For a given value of $q$ there exist four infinite sequences of characteristic values (eigenvalues) $a$, for either value of $a$ corresponding an infinite sequence (eigenvector) of expansion coefficients.

\subsection{Characteristic values and coefficients}

By subsituting (\ref{eq:8}) in (\ref{eq:6}), the following recurrence relations among the expansion coefficients result
\begin{eqnarray}
\label{eq:9}
&1 \quad \textrm{even-even:} \nonumber \\  
&aA_{ee}^{(0)}-qA_{ee}^{(2)}=0,  \nonumber \\ 
&(a-4)A_{ee}^{(2)}-q[2A_{ee}^{(0)}+A_{ee}^{(4)}]=0,  \nonumber \\
&[a-(2j)^2]A_{ee}^{(2j)}-q[A_{ee}^{(2j-2)}+A_{ee}^{(2j+2)}]=0, 
\qquad j=2,3,4\cdots  
\end{eqnarray} 
\begin{eqnarray}
\label{eq:10} 
&2 \quad \textrm{even-odd:} \nonumber \\
&(a-1)A_{eo}^{(1)}-q[A_{eo}^{(1)}+A_{eo}^{(3)}]=0, \nonumber \\
&[a-(2j+1)^2]A_{eo}^{(2j+1)}-q[A_{eo}^{(2j-1)}
+A_{eo}^{(2j+3)}]=0, \qquad j=1,2,3 \cdots 
\end{eqnarray} 
\begin{eqnarray}
\label{eq:11} 
&3 \quad \textrm{odd-even:} \nonumber  \\
&(a-4)A_{oe}^{(2)}-qA_{oe}^{(4)}=0,  \nonumber \\
&[a-(2j)^2]A_{oe}^{(2j)}-q[A_{oe}^{(2j-2)}+A_{oe}^{(2j+2)}]=0, 
\qquad j=2,3,4\cdots 
\end{eqnarray}
\begin{eqnarray}
\label{eq:12} 
&4 \quad \textrm{odd-odd:} \nonumber \\
&(a-1)A_{oo}^{(1)}+q[A_{oo}^{(1)}-A_{oo}^{(3)}]=0, \nonumber  \\
&[a-(2j+1)^2]A_{oo}^{(2j+1)}-q[A_{oo}^{(2j-1)}
+A_{oo}^{(2j+3)}]=0, 
\qquad j=1,2,3\cdots. 
\end{eqnarray}
The recurrence relations can be written in matrix form \cite{11},
\begin{eqnarray}
\label{eq:13}
1 \quad \textrm{even-even:} \qquad
\left(
\begin{array}{ccccccc}
-a & q & 0 & 0 & 0  & 0 & \cdots \\
2q &  2^2-a & q & 0 & 0 & 0 & \cdots \\
0 & q & 4^2-a & q & 0 & 0 & \cdots \\
0 & 0 & q & 6^2-a & q & 0 & \cdots \\
\vdots & \vdots & \vdots & \vdots & \vdots & \vdots & \ddots
\end{array} \right)
\left(
\begin{array}{c}
A_{ee}^{(0)} \\ A_{ee}^{(2)} \\ A_{ee}^{(4)} \\ A_{ee}^{(6)} \\
\vdots \end{array} \right) =0, 
\end{eqnarray}
\begin{eqnarray}
\label{eq:14}
2 \quad \textrm{even-odd:} \quad
\left(
\begin{array}{ccccccc}
1+q-a & q & 0 & 0 & 0  & 0 & \cdots \\
q &  3^2-a & q & 0 & 0 & 0 & \cdots \\
0 & q & 5^2-a & q & 0 & 0 & \cdots \\
0 & 0 & q & 7^2-a & q & 0 & \cdots \\
\vdots & \vdots & \vdots & \vdots & \vdots & \vdots & \ddots
\end{array} \right)
\left(
\begin{array}{c}
A_{eo}^{(1)} \\ A_{eo}^{(3)} \\ A_{eo}^{(5)} \\ A_{eo}^{(7)} \\
\vdots \end{array} \right)=0, 
\end{eqnarray}
\begin{eqnarray}
\label{eq:15}
3 \quad \textrm{odd-even:} \qquad
\left(
\begin{array}{ccccccc}
2^2-a & q & 0 & 0 & 0  & 0 & \cdots \\
q &  4^2-a & q & 0 & 0 & 0 & \cdots \\
0 & q & 6^2-a & q & 0 & 0 & \cdots \\
0 & 0 & q & 8^2-a & q & 0 & \cdots \\
\vdots & \vdots & \vdots & \vdots & \vdots & \vdots & \ddots
\end{array} \right)
\left(
\begin{array}{c}
A_{oe}^{(2)} \\ A_{oe}^{(4)} \\ A_{oe}^{(6)} \\ A_{oe}^{(8)} \\
\vdots \end{array} \right)=0, 
\end{eqnarray}
\begin{eqnarray}
\label{eq:16}
4 \quad \textrm{odd-odd:} \quad
\left(
\begin{array}{ccccccc}
1-q-a & q & 0 & 0 & 0  & 0 & \cdots \\
q &  3^2-a & q & 0 & 0 & 0 & \cdots \\
0 & q & 5^2-a & q & 0 & 0 & \cdots \\
0 & 0 & q & 7^2-a & q & 0 & \cdots \\
\vdots & \vdots & \vdots & \vdots & \vdots & \vdots & \ddots
\end{array} \right)
\left(
\begin{array}{c}
A_{oo}^{(1)} \\ A_{oo}^{(3)} \\ A_{oo}^{(5)} \\ A_{oo}^{(7)} \\
\vdots \end{array} \right)=0. 
\end{eqnarray}
The matrices are real, tridiagonal, and symmetric for all categories, with the exception of the ``1 even-even'' category where the matrix is slightly non-symmetric. The eigenvalue problem is accurately solved in Matlab. In other computer programs it could be necessary to transform the slightly non-symmetric matrix in a symmetric one \cite{11}. Both the eigenvalues $a$ and the corresponding eigenvectors ($A_{pm}$, with $p,m=e,o$) are determined for either category at any order $n$. The order $n$ takes different values for each category of Mathieu functions. For the purpose of avoiding any confusion, a distinction must be done between the $n^{th}$ order (in the succession of all orders) and the true value of that order. Thus, let denote $n$ the order in the succession of all orders, and $t$ the true value of order $n$. The values of $n$ and $t$ for the four categories of Mathieu functions are
\begin{eqnarray}
1 \qquad \textrm{even-even:} \qquad  
n=0,1,2\cdots \qquad t=0,2,4\cdots, \nonumber \\
2 \qquad \textrm{even-odd:} \qquad  
n=0,1,2\cdots \qquad t=1,3,5\cdots, \nonumber \\
3 \qquad \textrm{odd-even:} \qquad  
n=1,2,3\cdots \qquad t=2,4,6\cdots, \nonumber \\
4 \qquad \textrm{odd-odd:} \qquad  
n=0,1,2\cdots \qquad t=1,3,5\cdots.  \nonumber
\end{eqnarray}
Note that, if the notation is self-contained by all routines of Mathieu functions, there is no need to determine the specific values of $n$ and $t$ for either category of Mathieu functions since it is done automatically.

\subsection{Normalization and orthogonality}

Following \cite{6,11}, the angular Mathieu functions are normalized by requiring that
\begin{equation}
\label{eq:17}
S_{ep}(0,q,n)=1, \qquad 
\Big[\frac
{\mathrm{d} S_{op}(v,q,n)}{\mathrm{d} v} 
\Big]_{v=0}=1, 
\qquad p=e,o. 
\end{equation}
These requirements imply that,
\begin{eqnarray}
\label{eq:18} 
1 \qquad \textrm{even-even:}\qquad  \sum_{j=0}^\infty
A_{ee}^{(2j)}(q,n)=1, \quad \quad \qquad \nonumber \\
2 \qquad \textrm{even-odd:}\qquad  \sum_{j=0}^\infty
A_{eo}^{(2j+1)}(q,n)=1,\qquad \quad \\
3 \qquad \textrm{odd-even:}\qquad   \sum_{j=1}^\infty
2jA_{oe}^{(2j)}(q,n)=1,\qquad \quad \nonumber \\
4 \qquad \textrm{odd-odd:}\qquad  \sum_{j=0}^\infty
(2j+1)A_{oo}^{(2j+1)}(q,n)=1.  \nonumber
\end{eqnarray}
The orthogonality relation for the angular Mathieu functions is 
\begin{equation}
\label{eq:19}
\int_0^{2\pi}\! S_{pm}(v,q,n)S_{pm'}(v,q,n)\,
\mathrm{d} v=N_{pm}\delta_{m\,m'}, 
\qquad p,m,m'=e,o ,
\end{equation}
where $N_{pm}$ is normalization factor, $\delta_{m\,m'}$ equals 1 if $m=m'$ and equals 0 otherwise. Then, the following relations for the normalization factor result,
\begin{eqnarray}
\label{eq:20} 
1 \qquad \textrm{even-even:} \qquad  N_{ee}(q,n)=2\pi[A_{ee}^{(0)}(q,n)]^2+\pi\sum_{j=1}^\infty
[A_{ee}^{(2j)}(q,n)]^2, \nonumber \\
2 \qquad \textrm{even-odd:} \qquad 
N_{eo}(q,n)=\pi \sum_{j=0}^\infty
[A_{eo}^{(2j+1)}(q,n)]^2, \hspace{2.6cm} \\
3 \qquad \textrm{odd-even:}   \qquad  
N_{oe}(q,n)=\pi \sum_{j=1}^\infty
[A_{oe}^{(2j)}(q,n)]^2, \hspace{3cm} \nonumber \\
4 \qquad \textrm{odd-odd:} \qquad 
N_{oo}(q,n)=\pi \sum_{j=0}^\infty
[A_{oo}^{(2j+1)}(q,n)]^2. \hspace{2.7cm} \nonumber
\end{eqnarray}
Since different normalization schemes have been adopted in the literature, much attention should be paid when numerical results provided by different authors are compared ones against the others.

\subsection{Correlation factors}

Let consider two regions of different permittivities, $\epsilon$ and $\epsilon'$. The parameter $q$ being different in the two regions, $q \ne q'$, the characteristic values and expansion coefficients are also different. Let $S_{pm}$ and $S_{pm}^\prime$ be the respective angular Mathieu functions. The correlation factors $C_{pm}(q,q',n)$, with $p,m=e,o$, between the angular Mathieu functions $S_{pm}$ and $S_{pm}^\prime$ are defined by relation
\begin{equation}
\label{eq:21}
C_{pm}(q,q',n)=\delta_{m\,m'} \int_0^{2\pi} S_{pm'}(v,q,n)S_{pm}^\prime(v,q',n)\,\mathrm{d} v ,  
\qquad  p,m,m'=e,o. 
\end{equation} 
Using (\ref{eq:8}) gives
\begin{eqnarray}
\label{eq:22}
1  \quad \textrm{even-even:} \qquad
C_{ee}(q,q',n)=2\pi A_{ee}^{(0)}(q,n)A_{ee}^{\prime \,(0)}(q',n)\hspace{2cm} \nonumber \\
  +\pi\sum_{j=1}^\infty
A_{ee}^{(2j)}(q,n)A_{ee}^{\prime \,(2j)}(q',n), \hspace{1cm} \nonumber \\
2 \qquad \textrm{even-odd:} \qquad
C_{eo}(q,q',n)=\pi \sum_{j=0}^\infty A_{eo}^{(2j+1)}(q,n)A_{eo}^{\prime \,
(2j+1)}(q',n), \\
3 \qquad \textrm{odd-even:} \qquad 
C_{oe}(q,q',n)=\pi \sum_{j=1}^\infty
A_{oe}^{(2j)}(q,n)A_{oe}^{\prime \,(2j)}(q',n), \hspace{0.7cm} \nonumber \\
4 \qquad \textrm{odd-odd:} \qquad
C_{oo}(q,q',n)=\pi \sum_{j=0}^\infty
A_{oo}^{(2j+1)}(q,n)A_{oo}^{\prime \,(2j+1)}(q',n).\hspace{0.1cm} \nonumber
\end{eqnarray}

\subsection{Derivatives of angular Mathieu functions}

The derivatives of the angular Mathieu functions follow readily from (\ref{eq:8}),
\begin{eqnarray}
\label{eq:23} 
1 \qquad \textrm{even-even:}\qquad  
\frac{\mathrm{d} S_{ee}(v,q,n)}{\mathrm{d} v}=-\sum_{j=1}^\infty
2jA_{ee}^{(2j)}(q,n)\sin(2jv), \hspace{2.1cm} \nonumber \\
2 \qquad \textrm{even-odd:}\qquad  
\frac{\mathrm{d} S_{eo}(v,q,n)}{\mathrm{d} v}=-\sum_{j=0}^\infty
(2j+1)A_{eo}^{(2j+1)}(q,n)\sin[(2j+1)v], \\
3 \qquad \textrm{odd-even:}\qquad   
\frac{\mathrm{d} S_{oe}(v,q,n)}{\mathrm{d} v}=\sum_{j=1}^\infty
2jA_{oe}^{(2j)}(q,n)\cos(2jv),\hspace{2.6cm} \nonumber \\
4 \qquad \textrm{odd-odd:}\qquad  
\frac{\mathrm{d} S_{oo}(v,q,n)}{\mathrm{d} v}=\sum_{j=0}^\infty
(2j+1)A_{oo}^{(2j+1)}(q,n)\cos[(2j+1)v]. \hspace{0.5cm} \nonumber
\end{eqnarray}

\section{Radial Mathieu functions}

Solutions of (\ref{eq:7}) can be obtained from (\ref{eq:8}) by replacing $v$ by $iu$. Instead of $\sin v$ and $\cos v$, the terms of the series now involve $\sinh u$ and $\cosh u$. The convergence is low unless $|u|$ is small. Better convergence of series results by expressing the solutions of (\ref{eq:7}) in terms of Bessel functions associated with the same expansion coefficients that are determined once for both the angular and radial Mathieu functions. Either pair of angular and radial Mathieu functions are proportional to one another \cite{6},
\begin{equation}
\label{eq:24}
S_{ep}(iu,q,n)=\sqrt{2\pi}g_{ep}(q,n)J_{ep}(u,q,n), \qquad p=e,o, 
\end{equation}
where $J_{ep}$ are even radial Mathieu functions of the first kind and $g_{ep}$ are joining factors. When $u=0$,
\begin{equation}
\label{eq:25}
S_{ep}(0,q,n)=1, \qquad  J_{ep}(0,q,n)=\frac{1}{\sqrt{2\pi}g_{ep}(q,n)}, \qquad p=e,o .
\end{equation}
Thus, one obtains,
\begin{align}
\label{eq:26}
&1 \quad \textrm{even-even:} \qquad 
g_{ee}(q,n)=\frac{(-1)^r}{\pi A_{ee}^{(0)}(q,n)}S_{ee}(\pi/2,q,n), 
\qquad  r=t/2, \nonumber \\
&2 \quad \textrm{even-odd:} \qquad
g_{eo}(q,n)=\frac{-(-1)^r}{\pi \sqrt{q} A_{eo}^{(1)}(q,n)}\Big[\frac{\mathrm{d} S_{eo}(v,q,n)} 
{\mathrm{d} v}\Big]_{v=\pi/2}, 
\qquad r=(t-1)/2. 
\end{align}
Similarly \cite{6},
\begin{equation}
\label{eq:27}
-iS_{op}(iu,q,n)=\sqrt{2\pi}g_{op}(q,n)J_{op}(u,q,n), \qquad p=e,o. 
\end{equation}
When $u=0$,
\begin{equation}
\label{eq:28}
J_{op}(0,q,n)=0, \qquad  \Big[ 
\frac{\mathrm{d} J_{op}(u,q,n)}{\mathrm{d} u}
\Big]_{u=0}=\frac{1}{\sqrt{2\pi}g_{op}(q,n)},\qquad  
p=e,o. 
\end{equation}
Thus, one obtains,
\begin{eqnarray}
\label{eq:29}
3 \quad  \textrm{odd-even:} \qquad 
g_{oe}(q,n)=\frac{(-1)^r}{\pi q A_{oe}^{(2)}(q,n)}\Big[\frac{\mathrm{d} S_{oe}(v,q,n)}
{\mathrm{d} v}\Big]_{v=\pi/2},\qquad  r=t/2, \nonumber \\
4 \quad  \textrm{odd-odd:} \qquad 
g_{oo}(q,n)=\frac{(-1)^r}{\pi \sqrt{q} A_{oo}^{(1)}(q,n)}S_{oo}(\pi/2,q,n),\qquad  r=(t-1)/2. 
\end{eqnarray}
Remember that $t$ is the true value of order $n$.

\subsection{Radial Mathieu functions of the first kind}

Since rapidly converging series are those expressed in terms of products of Bessel functions \cite{10,11}, in the following relations refer only to them. Similarly to the angular Mathieu functions, one may distinct four categories of radial Mathieu functions of the first kind which are denoted $J_{pm}(u,q,n)$, with $p,m=e,o$,
\begin{align}
\label{eq:30} 
&1\quad \textrm{even-even:}  
 \qquad J_{ee}(u,q,n) =\sqrt{\frac{\pi}{2}}\frac{(-1)^r}
{A_{ee}^{(0)}(q,n)}
\sum_{j=0}^\infty(-1)^jA_{ee}^{(2j)}(q,n)J_j(v_1)J_j(v_2), \nonumber \\
&\hspace{5cm} r=t/2,\nonumber \\
&2\quad \textrm{even-odd:} 
 \qquad J_{eo}(u,q,n) =\sqrt{\frac{\pi}{2}}\frac{(-1)^r}
{A_{eo}^{(1)}(q,n)}
\sum_{j=0}^\infty(-1)^jA_{eo}^{(2j+1)}(q,n)[J_j(v_1)J_{j+1}(v_2)\nonumber \\ 
& \hspace{4cm} +J_j(v_2)J_{j+1}(v_1)], \qquad r=(t-1)/2, \\
&3\quad\textrm{odd-even:} \qquad
J_{oe}(u,q,n)=\sqrt{\frac{\pi}{2}}\frac{(-1)^r}
{A_{oe}^{(2)}(q,n)}
\sum_{j=1}^\infty (-1)^j A_{oe}^{(2j)}(q,n)[J_{j-1}(v_1) J_{j+1}(v_2)
 \nonumber \\
& \hspace{4cm}-J_{j-1}(v_2)J_{j+1}(v_1)], \qquad r=t/2,\nonumber \\
&4\quad\textrm{odd-odd:} \qquad
J_{oo}(u,q,n)=\sqrt{\frac{\pi}{2}}\frac{(-1)^r}{A_{oo}^{(1)}(q,n)}
\sum_{j=0}^\infty(-1)^j A_{oo}^{(2j+1)}(q,n)[J_j(v_1)J_{j+1}(v_2) \nonumber \\
&\hspace{5cm}-J_j(v_2)J_{j+1}(v_1)], \qquad r=(t-1)/2,  \nonumber
\end{align}
where $v_1=\sqrt{q}\exp{(-u)}$ and $v_2=\sqrt{q}\exp{(u)}$. The derivatives of the radial Mathieu functions of the first kind are
\flushleft 1 \hspace{0.5cm} even-even:$\qquad r=t/2,$ 
\begin{align}
\frac{\mathrm{d} J_{ee}(u,q,n)}{\mathrm{d} u}
=&\sqrt{\frac{\pi}{2}}\frac{(-1)^r}{A_{ee}^{(0)}(q,n)} 
\sum_{j=0}^\infty(-1)^j A_{ee}^{(2j)}(q,n) 
[v_1J_{j+1}(v_1)J_j(v_2) \nonumber \\
&-v_2J_j(v_1)J_{j+1}(v_2)],  \nonumber
\end{align}
\flushleft 2 \hspace{0.5cm} even-odd:$\qquad r=(t-1)/2,$ 
\begin{align}   
\frac{\mathrm{d} J_{eo}(u,q,n)}{\mathrm{d} u}=&   \sqrt{\frac{\pi}{2}}\frac{(-1)^r}{A_{eo}^{(1)}(q,n)}
\sum_{j=0}^\infty(-1)^j A_{eo}^{(2j+1)}(q,n)  
\Big\{(v_2-v_1)[J_j(v_1)J_j(v_2) \nonumber \\
&-J_{j+1}(v_1)J_{j+1}(v_2)]  
+ (2j+1)[J_{j+1}(v_1)J_j(v_2)-J_j(v_1)J_{j+1}(v_2)]\Big\},
 \nonumber
\end{align}
\flushleft 3 \hspace{0.5cm} odd-even:$\qquad r=t/2,$ 
\begin{align}
\label{eq:31}
\frac{\mathrm{d} J_{oe}(u,q,n)}{\mathrm{d} u}=&  \sqrt{\frac{\pi}{2}}\frac{(-1)^r}{A_{oe}^{(2)}(q,n)}
\sum_{j=0}^\infty(-1)^{j+1} A_{oe}^{(2j+2)}(q,n)(4j+4)  
\Big\{J_j(v_1)J_j(v_2)  \\
&+\cosh 2u J_{j+1}(v_1)J_{j+1}(v_2)  
-(j+1)[\frac{1}{v_1}J_{j+1}(v_1)J_j(v_2)
+\frac{1}{v_2}J_j(v_1)J_{j+1}(v_2)]
\Big\},  \nonumber 
\end{align}
\flushleft 4 \hspace{0.5cm} odd-odd:$\qquad r=(t-1)/2,$ 
\begin{align}
\frac{\mathrm{d} J_{oo}(u,q,n)}{\mathrm{d} u}=&   \sqrt{\frac{\pi}{2}}\frac{(-1)^r}{A_{oo}^{(1)}(q,n)}
\sum_{j=0}^\infty(-1)^{j} A_{oo}^{(2j+1)}(q,n)  
\Big\{(v_1+v_2)[J_j(v_1)J_j(v_2)  \nonumber \\
&+J_{j+1}(v_1)J_{j+1}(v_2)]  
- (2j+1)[J_{j+1}(v_1)J_j(v_2)+J_j(v_1)J_{j+1}(v_2)]\Big\}.  
 \nonumber
\end{align}  

\subsection{Radial Mathieu functions of the second kind}

A second independent solution of (\ref{eq:7}) is obtained by replacing the Bessel functions of the first kind $J_n(v_2)$ in (\ref{eq:30}) by the Bessel functions of the second kind $Y_n(v_2)$ \cite{10,11}. This solution is denoted $Y_{pm}(u,q,n)$, with $p,m=e,o$.

\begin{align}
\label{eq:32} 
&1\quad \textrm{even-even:}  
 \qquad Y_{ee}(u,q,n) =\sqrt{\frac{\pi}{2}}\frac{(-1)^r}
{A_{ee}^{(0)}(q,n)}
\sum_{j=0}^\infty(-1)^jA_{ee}^{(2j)}(q,n)J_j(v_1)Y_j(v_2), \nonumber \\
&\hspace{5cm} r=t/2,\nonumber \\
&2\quad \textrm{even-odd:} 
 \qquad Y_{eo}(u,q,n) =\sqrt{\frac{\pi}{2}}\frac{(-1)^r}
{A_{eo}^{(1)}(q,n)}
\sum_{j=0}^\infty(-1)^jA_{eo}^{(2j+1)}(q,n)[J_j(v_1)Y_{j+1}(v_2)\nonumber \\ 
& \hspace{4cm} +Y_j(v_2)J_{j+1}(v_1)], \qquad r=(t-1)/2, \\
&3\quad\textrm{odd-even:} \qquad
Y_{oe}(u,q,n)=\sqrt{\frac{\pi}{2}}\frac{(-1)^r}
{A_{oe}^{(2)}(q,n)}
\sum_{j=1}^\infty (-1)^j A_{oe}^{(2j)}(q,n)[J_{j-1}(v_1) Y_{j+1}(v_2)
 \nonumber \\
& \hspace{4cm}-Y_{j-1}(v_2)J_{j+1}(v_1)], \qquad r=t/2,\nonumber \\
&4\quad\textrm{odd-odd:} \qquad
Y_{oo}(u,q,n)=\sqrt{\frac{\pi}{2}}\frac{(-1)^r}{A_{oo}^{(1)}(q,n)}
\sum_{j=0}^\infty(-1)^j A_{oo}^{(2j+1)}(q,n)[J_j(v_1)Y_{j+1}(v_2) \nonumber \\
&\hspace{5cm}-Y_j(v_2)J_{j+1}(v_1)], \qquad r=(t-1)/2,  \nonumber
\end{align}
The derivatives of the radial Mathieu functions of the second kind are
\flushleft 1 \hspace{0.5cm} even-even:$\qquad r=t/2,$ 
\begin{align}
\frac{\mathrm{d} Y_{ee}(u,q,n)}{\mathrm{d} u}
=&\sqrt{\frac{\pi}{2}}\frac{(-1)^r}{A_{ee}^{(0)}(q,n)} 
\sum_{j=0}^\infty(-1)^j A_{ee}^{(2j)}(q,n) 
[v_1J_{j+1}(v_1)Y_j(v_2) \nonumber \\
&-v_2J_j(v_1)Y_{j+1}(v_2)],  \nonumber
\end{align}
\flushleft 2 \hspace{0.5cm} even-odd:$\qquad r=(t-1)/2,$ 
\begin{align}   
\frac{\mathrm{d} Y_{eo}(u,q,n)}{\mathrm{d} u}=&   \sqrt{\frac{\pi}{2}}\frac{(-1)^r}{A_{eo}^{(1)}(q,n)}
\sum_{j=0}^\infty(-1)^j A_{eo}^{(2j+1)}(q,n)  
\Big\{(v_2-v_1)[J_j(v_1)Y_j(v_2) \nonumber \\
&-J_{j+1}(v_1)Y_{j+1}(v_2)]  
+ (2j+1)[J_{j+1}(v_1)Y_j(v_2)-J_j(v_1)Y_{j+1}(v_2)]\Big\},
 \nonumber
\end{align}
\flushleft 3 \hspace{0.5cm} odd-even:$\qquad r=t/2,$ 
\begin{align}
\label{eq:33}
\frac{\mathrm{d} Y_{oe}(u,q,n)}{\mathrm{d} u}=&  \sqrt{\frac{\pi}{2}}\frac{(-1)^r}{A_{oe}^{(2)}(q,n)}
\sum_{j=0}^\infty(-1)^{j+1} A_{oe}^{(2j+2)}(q,n)(4j+4)  
\Big\{J_j(v_1)Y_j(v_2)  \\
&+\cosh 2u J_{j+1}(v_1)Y_{j+1}(v_2)  
-(j+1)[\frac{1}{v_1}J_{j+1}(v_1)Y_j(v_2)
+\frac{1}{v_2}J_j(v_1)Y_{j+1}(v_2)]
\Big\},  \nonumber 
\end{align}
\flushleft 4 \hspace{0.5cm} odd-odd:$\qquad r=(t-1)/2,$ 
\begin{align}
\frac{\mathrm{d} Y_{oo}(u,q,n)}{\mathrm{d} u}=&   \sqrt{\frac{\pi}{2}}\frac{(-1)^r}{A_{oo}^{(1)}(q,n)}
\sum_{j=0}^\infty(-1)^{j} A_{oo}^{(2j+1)}(q,n)  
\Big\{(v_1+v_2)[J_j(v_1)Y_j(v_2)  \nonumber \\
&+J_{j+1}(v_1)Y_{j+1}(v_2)]  
- (2j+1)[J_{j+1}(v_1)Y_j(v_2)+J_j(v_1)Y_{j+1}(v_2)]\Big\}.  
 \nonumber
\end{align}  

\subsection{Radial Mathieu functions of the third and the fourth kinds}

Radial Mathieu functions of the third kind, analogous to the Hankel functions of the first kind are defined as follows \cite{6,11}
\begin{equation}
\label{eq:34}
H_{pm1}(u,q,n)=J_{pm}(u,q,n)+iY_{pm}(u,q,n), 
\qquad p,m=e,o. 
\end{equation} 
Similarly, radial Mathieu functions of the fourth kind, analogous to the Hankel functions of the second kind are defined as follows \cite{6,11}
\begin{equation}
\label{eq:35}
H_{pm2}(u,q,n)=J_{pm}(u,q,n)-iY_{pm}(u,q,n),
\qquad p,m=e,o.
\end{equation} 

\section{Implementation of Mathieu functions in Matlab}

Following the notation of the four categories of angular Mathieu functions, the implementation in Matlab or in any other computer program is readily done by introducing a function code $KF$. The first step in any algorithm of Mathieu function computation is to find the characteristic values (eigenvalues) and the expansion coefficients (eigenvectors). In \cite{13}, this is done by routine ``eig\_Spm'' which has $q$ as input parameter (see Table~\ref{tab:1}). Besides $q$, the function code $KF$ should be specified. Thus, if $KF=1$, the routine ``eig\_Spm'' solves the eigenvalue problem for category ``1 even-even'' of Mathieu functions, if $KF=2$ for category ``2 even-odd'', and so on. The number of expansion coefficients is the same, it is set equal to 25, for all categories of Mathieu functions. Concerning the outputs of routine ``eig\_Spm'', $va$ is a line vector representing the characteristic values $a$ for all the 25 orders; $mc$ is $25\times25$ matrix, where the columns represent the eigenvectors (that is, the expansion coefficients) for all orders; $vt$ is a column vector specifying the true value $t$ for all orders. Note that the eigenvectors in $mc$ were processed to obey equation~(\ref{eq:18}).
For the purpose to save the time of computation, all the other routines have $mc$ as input (see Table~\ref{tab:1}), the routine ``eig\_Spm'' being called once, at the beginning of the computation, for any values of coordinates $u$ and $v$ that intervene in that computation. Since in many cases the convergence is assured by the first several orders, all the other routines have $nmax\leq25$ as input. It means that those routines take into account only the first $nmax$ orders, but for either order the length of the corresponding eigenvector is the same, equal to 25. The routine ``extract\_one\_value'' can be used to get a single value, and the routine ``extract\_one\_column'' to get a single eigenvector, corresponding to the order $t$. The derivatives of $S_{pm}$, with $p,m=e,o$, are computed by routine ``dSpm''. For both ``Spm'' and ``dSpm'', $v$ is expressed in radians, with values in interval $(0,2\pi)$. The normalization, correlation, and joining factors are computed by routines ``Npm'', ``Cpm'', and ``gpm'', respectively. The four kinds of radial Mathieu functions, $J_{pm},Y_{pm},H_{pm1},~\textrm{and} ~H_{pm2}$, with $p,m=e,o$, are computed by routines ``Jpm'',``Ypm'',``Hpm1'', and ``Hpm2'', respectively, and their derivatives with respect to $u$ by routines ``dJpm'',``dYpm'',``dHpm1'', and ``dHpm2'', respectively. 

Numerical values of the separation constant $a$, of the angular Mathieu functions $S_{pm}$ and their derivatives $S_{pm}^\prime$, with $p,m=e,o$, where the prime denotes differentiation with respect to $v$, are given in Tables~\ref{tab:2}--\ref{tab:4}. They can be compared with data in \cite{2}. With the purpose to facilitate the comparison, since in \cite{2} the normalization $N_{pm}=\pi$ is applied, the data of $S_{pm}$ and $S_{pm}^\prime$ in Tables~\ref{tab:2}--\ref{tab:4} are multiplied by $\sqrt{\pi/N_{pm}}$.

Concerning the radial Mathieu functions, numerical values of $S_{ep}(iu,q,n)$ and $-iS_{op}(iu,q,n)$ are given for $u=0.5$ in Tables~\ref{tab:5} and~\ref{tab:6}. They are multiplied by $\sqrt{\pi/N_{pm}}$ and compared with data in \cite{9}. Note that $S_{ep}$ is correlated to the radial Mathieu function of the first kind $J_{ep}$ by Eq.~(\ref{eq:24}), whereas $S_{op}$ is correlated to $J_{op}$ by Eq.~(\ref{eq:27}). We found that, for parameters in \cite{9}, the values of $S_{ep}(iu,q,n)$ and $-iS_{op}(iu,q,n)$ calculated with Eqs.~(\ref{eq:24}) and~(\ref{eq:27}) differ from those obtained with Eq.~(\ref{eq:8}) by less than $7.5\times10^{-12}$.

\begin{table}[ht]
\begin{center}
\caption{\label{tab:1}Routines comprised in the toolbox \cite{13}.}
\begin{tabular}{|l|l|l|}
\hline
Name of routine & Routine call & What the routine computes \\
\hline          
eig\_Spm&$[va,mc,vt]$=eig\_Spm$(KF,q)$ &Vector of characteristic values $va$, matrix of  \\&&coefficients $mc$, and vector of orders $vt$, at given \\
&&function code $KF$ and elliptical parameter $q \ge 0$. \\ 
Spm&$y$=Spm$(KF,v,mc,nmax)$&Angular Mathieu functions $S_{pm}$, 
[Eq.~(\ref{eq:8})]. \\
dSpm&$y$=dSpm$(KF,v,mc,nmax)$&Derivatives with respect to $v$ of $S_{pm}$, 
[Eq.~(\ref{eq:23})]. \\
Npm&$y$=Npm$(KF,mc,nmax)$&Normalizing factors of angular Mathieu functions \\
&&$S_{pm}$, [Eqs.~(\ref{eq:19}) and~(\ref{eq:20})]. \\
Cpm&$y$=Cpm$(KF,mc,mc',nmax)$&Correlation factors of $S_{pm}$ and $S_{pm}^\prime$, having matrices \\
&& of coefficients $mc$ and $mc'$,  
[Eqs.~(\ref{eq:21}) and~(\ref{eq:22})]. \\
Jpm&$y$=Jpm$(KF,u,q,mc,nmax)$&Radial Mathieu functions of the first kind $J_{pm}$, \\&&[Eq.~(\ref{eq:30})]. \\
dJpm&$y$=dJpm$(KF,u,q,mc,nmax)$&Derivatives with respect to $u$ of $J_{pm}$, 
[Eq.~(\ref{eq:31})]. \\
gpm&$y$=gpm$(KF,q,mc,nmax)$&Joining factors for pairs of angular, $S_{pm}$ and \\
&&radial, $J_{pm}$ Mathieu functions, 
[Eqs.~(\ref{eq:24})--(\ref{eq:29})]. \\
Ypm&$y$=Ypm$(KF,u,q,mc,nmax)$&Radial Mathieu functions of the second kind \\
 &&$Y_{pm}$, [Eq.~(\ref{eq:32})]. \\
dYpm&$y$=dYpm$(KF,u,q,mc,nmax)$&Derivatives with respect to $u$ of $Y_{pm}$, 
[Eq.~(\ref{eq:33})]. \\
Hpm1&$y$=Hpm1$(KF,u,q,mc,nmax)$&Radial Mathieu functions of the third kind 
$H_{pm1}$, \\&& [Eq.~(\ref{eq:34})]. \\
dHpm1&$y$=dHpm1$(KF,u,q,mc,namax)$&Derivatives with respect to $u$ of $H_{pm1}$.  \\
Hpm2&$y$=Hpm2$(KF,u,q,mc,nmax)$&Radial Mathieu functions of the fourth kind 
$H_{pm2}$, \\&& [Eq.~(\ref{eq:35})]. \\
dHpm2&$y$=dHpm2$(KF,u,q,mc,namax)$&Derivatives with respect to $u$ of $H_{pm2}$. \\
extract\_one\_column&$y$=extract\_one\_column$(KF,t,mc)$&Extracts one column from $mc$ at given $t$. \\
extract\_one\_value&$y$=extract\_one\_value$(KF,t,vec)$&Extracts one value from $vec$ at given $t$. \\
\hline
\end{tabular}
\end{center}
\end{table}
\begin{table}[ht]
\begin{center}
\caption{\label{tab:2}Values of $S_{ee}$ 
multiplied by $\gamma_{ee}=\sqrt{\pi/N_{ee}}$ to be compared with data in \cite{2}}
\begin{tabular}{|c|c|c|c|c|}
\hline
t & q & $a$ & $\gamma_{ee}S_{ee}(0,q,n)$ & $\gamma_{ee}S_{ee}(\pi/2,q,n)$\\
\hline          
  0 &   0 &                 0 &  0.7071067811865 &  0.7071067811865 \\
    &   5 &  -5.8000460208515 &  0.0448001816519 &  1.3348486746980 \\
    &  10 & -13.9369799566589 &  0.0076265175709 &  1.4686604707129 \\
    &  15 & -22.5130377608640 &  0.0019325083152 &  1.5501081466866 \\
    &  20 & -31.3133900703364 &  0.0006037438292 &  1.6098908573959 \\
    &  25 & -40.2567795465667 &  0.0002158630184 &  1.6575102983235 \\
\hline
   2 &  0 & 4.0000000000000 &  1.0000000000000 & -1.0000000000000 \\
     &  5 & 7.4491097395292 &  0.7352943084007 & -0.7244881519677 \\
     & 10 & 7.7173698497796 &  0.2458883492913 & -0.9267592641263 \\
     & 15 & 5.0779831975435 &  0.0787928278464 & -1.0199662260303 \\
     & 20 & 1.1542828852468 &  0.0286489431471 & -1.0752932287797 \\
     & 25 &-3.5221647271583 &  0.0115128663309 & -1.1162789532953 \\
\hline
10 &  0  &   100.0000000000000 &  1.0000000000000 & -1.0000000000000 \\
   &  5  &   100.1263692161636 &  1.0259950270894 & -0.9753474872360 \\
   & 10  &   100.5067700246816 &  1.0538159921009 & -0.9516453181790 \\
   & 15  &   101.1452034473016 &  1.0841063118392 & -0.9285480638845 \\
   & 20  &   102.0489160244372 &  1.1177886312594 & -0.9057107845941 \\ 
   & 25  &   103.2302048044949 &  1.1562399186322 & -0.8826919105637 \\
\hline
\end{tabular}
\end{center}
\end{table}
\begin{table}[ht]
\begin{center}
\caption{\label{tab:3}Values of $S_{eo}$ and $S_{eo}^\prime$
multiplied by $\gamma_{eo}=\sqrt{\pi /N_{eo}}$ to be compared with data in \cite{2}}
\begin{tabular}{|c|c|c|c|c|}
\hline
t & q & $a$ & $\gamma_{eo}S_{eo}(0,q,n)$ & $\gamma_{eo}S_{eo}^\prime(\pi/2,q,n)$ \\
\hline
  1 & 0  & 1.0000000000000 & 1.0000000000000 & -1.0000000000000 \\
    & 5  & 1.8581875415478 & 0.2565428793224 & -3.4690420034057 \\
    &10  &-2.3991424000363 & 0.0535987477472 & -4.8504383044964 \\
    &15  &-8.1011051316418 & 0.0150400664538 & -5.7642064390510 \\
    &20 &-14.4913014251748 & 0.0050518137647 & -6.4905657825800 \\
    &25 &-21.3148996906657 & 0.0019110515067 & -7.1067412352901 \\
\hline                    
5  &  0   &    25.0000000000000 &  1.0000000000000 & -5.0000000000000 \\
   &  5   &    25.5499717499816 &  1.1248072506385 & -5.3924861549882 \\
   & 10   &    27.7037687339393 &  1.2580199413083 & -5.3212765411609 \\
   & 15   &    31.9578212521729 &  1.1934322304131 & -5.1191498884064 \\
   & 20   &    36.6449897341328 &  0.9365755314226 & -5.7786752500644 \\
   & 25   &    40.0501909858077 &  0.6106943100507 & -7.0598842916553 \\
 \hline                    
  15 &  0 &   225.0000000000000 &  1.0000000000000 &  15.0000000000000 \\
     &  5 &   225.0558124767096 &  1.0112937325296 &  15.1636574720602 \\
     & 10 &   225.2233569749644 &  1.0228782824382 &  15.3198803056623 \\
     & 15 &   225.5029562446541 &  1.0347936522369 &  15.4687435032830 \\
     & 20 &   225.8951534162079 &  1.0470843441629 &  15.6102785232380 \\
     & 25 &   226.4007200447481 &  1.0598004418139 &  15.7444725050679 \\ 
\hline
\end{tabular}
\end{center}
\end{table}
\begin{table}[ht]
\begin{center}
\caption{\label{tab:4}Values of $S_{op}$ and $S_{op}^\prime$ multiplied by $\sqrt{\pi /N_{op}}, p=e,o$ (see~\cite{2})}
\begin{tabular}{|c|c|c|c|c|}
\hline
t & q & $a$ & $\sqrt{\pi /N_{oe}}S_{oe}^\prime(0,q,n)$ & $\sqrt{\pi /N_{oe}}S_{oe}^\prime(\pi/2,q,n)$\\
\hline
  2 &   0 &    4.0000000000000 &  2.0000000000000 & -2.0000000000000 \\
    &   5 &    2.0994604454867 &  0.7331661960372 & -3.6405178524082 \\
    &  10 &   -2.3821582359570 &  0.2488228403985 & -4.8634220691653 \\
    &  15 &   -8.0993467988959 &  0.0918197143696 & -5.7655737717278 \\
    &  20 &  -14.4910632559807 &  0.0370277776852 & -6.4907522240373 \\
    &  25 &  -21.3148606222498 &  0.0160562170491 & -7.1067719073739 \\        
 10 &   0 &  100.0000000000000 & 10.0000000000000 & -10.0000000000000 \\
    &   5 &  100.1263692156019  & 9.7341731518695 & -10.2396462566908 \\
    &  10 &  100.5067694628784  & 9.4404054347686 & -10.4539475316485 \\
    &  15 &  101.1451722929092  & 9.1157513395126 & -10.6428998776563 \\
    &  20 &  102.0483928609361  & 8.7555450801360 & -10.8057241781325 \\
    &  25 &  103.2256800423735  & 8.3526783655914 & -10.9413538308191 \\
\hline
t & q & $a$ & $\sqrt{\pi /N_{oo}}S_{oo}^\prime(0,q,n)$ & $\sqrt{\pi /N_{oo}}S_{oo}(\pi/2,q,n)$\\
\hline
   1 &    0  &  1.0000000000000 &  1.0000000000000 &  1.0000000000000 \\
     &    5  & -5.7900805986378 &  0.1746754006198 &  1.3374338870223 \\
     &   10 & -13.9365524792501 &  0.0440225659111 &  1.4687556641029 \\
     &   15 & -22.5130034974235 &  0.0139251347875 &  1.5501150743576 \\
     &   20 & -31.3133861669129 &  0.0050778849001 &  1.6098915926038 \\
     &   25 & -40.2567789846842 &  0.0020443593656 &  1.6575103983745 \\        
   5 &    0 &  25.0000000000000 &  5.0000000000000 & 1.0000000000000 \\
     &    5 &  25.5108160463032 &  4.3395700104946 & 0.9060779302024 \\
     &   10 &  26.7664263604801 &  3.4072267604013 & 0.8460384335355 \\
     &   15 &  27.9678805967175 &  2.4116664728002 & 0.8379493400125 \\
     &   20 &  28.4682213251027 &  1.5688968684857 & 0.8635431218534 \\
     &   25 &  28.0627658994543 &  0.9640716219024 & 0.8992683245108 \\         
  15 &    0 & 225.0000000000000 &  15.0000000000000 & -1.0000000000000 \\
     &    5 & 225.0558124767096 &  14.8287889732852 & -0.9889607027406 \\
     &   10 & 225.2233569749643 &  14.6498600449581 & -0.9781423471832 \\
     &   15 & 225.5029562446537 &  14.4630006940372 & -0.9675137031855 \\
     &   20 & 225.8951534161767 &  14.2679460909928 & -0.9570452540613 \\ 
     &   25 & 226.4007200438825 &  14.0643732956172 & -0.9467086958781 \\
\hline
\end{tabular}
\end{center}
\end{table}
\begin{table}[ht]
\begin{center}
\caption{\label{tab:5}Values of $S_{ep}(iu,q,n)$ for $u=0.5$ multiplied by $\sqrt{\pi/N_{ep}}$, where $p=e,o$, compared with data in \cite{9}}
\begin{tabular}{|c|c|c|c||c|c|c|c|}
\hline
t&q& Values at $p=e$&Data in \cite{9}&t&q&Values at $p=o$&Data in \cite{9}\\
\hline
0&    5 & -0.019325304910071& -0.01932&1& 5& 0.021440743185527&  0.02144\\
 &   10 & -0.007055239716193& -0.00705& &10&-0.038634237458525& -0.03863  \\
 &   20 & -0.000169411415735& -0.00016& &20&-0.003373888309642& -0.00337   \\
2&    5 &  0.446937465741068&  0.44693&3& 5& 1.205528267066838&  1.2055  \\
 &   10 & -0.063855921612085& -0.06385& &10& 0.235940782144547&  0.23594   \\
 &   20 & -0.024916657795101& -0.02491& &20&-0.097385461808731& -0.09738  \\
4&    5 &  2.234088244534832&  2.2341 &5& 5& 3.864089377116713&  3.8641 \\
 &   10 &  1.039103163573830&  1.0391 & &10& 2.285610444240526&  2.2856  \\
 &   20 & -0.143991090269732& -0.14399& &20& 0.274270780278172&  0.27427  \\
\hline 
\end{tabular}
\end{center}
\end{table}
\begin{table}[h]
\begin{center}
\caption{\label{tab:6}Values of $-iS_{op}(iu,q,n)$ for $u=0.5$ multiplied by $\sqrt{\pi /N_{op}}$, with $p=e,o$, compared with data in \cite{9}}
\begin{tabular}{|c|c|c|c||c|c|c|c|}
\hline
t&q&Values at $p=e$&Data in \cite{9}&t&q&Values at $p=o$&Data in \cite{9}\\
\hline
2& 5 & 0.238342768735937 & 0.23834 &1& 5& 0.036613617783886 &  0.03661  \\
 &10 & 0.028675814044625 & 0.02867 & &10& 0.000750806874015 &  0.00075 \\  
 &20 &-0.003176296415956 &-0.00317 & &20&-0.000538258353937 & -0.00053  \\
4& 5 & 1.883560277440876 & 1.8836  &3& 5& 0.806555153528872 &  0.80655 \\
 &10 & 0.769679129538722 & 0.76968 & &10& 0.204495885546638 &  0.20449 \\
 &20 & 0.040515136278697 &  0.04051& &20&-0.005279473480675 & -0.00527  \\
6& 5 & 6.6066602369876   &  6.6067 &5& 5& 3.667530204538722 & 3.6675 \\ 
 &10 & 4.1161420952367   &  4.1161 & &10& 1.972361938552091 & 1.9724 \\
 &20 & 1.1805904286267   &  1.1806 & &20& 0.320398855944192 & 0.32040 \\ 
\hline
\end{tabular}
\end{center}
\end{table}


\begin{thebibliography}{15}

\bibitem{1}E. Mathieu ``Le mouvement vibratoire d'une membrane de forme elliptique,'' Jour. de Math. Pures at Appliquees (Jour. de Liouville) {\bf 13}, 137--203 (1868).

\bibitem{2}M. Abramowitz and I. Stegun {\it Handbook of Mathematical Functions} (New York, 1964).

\bibitem{3}I. S. Gradshteyn and I. M. Ryzhik {\it Tables of Integrals, Series, and Products} (Academic Press, San Diego, 1994).

\bibitem{4}E. L. Ince {\it Ordinary Differential Equations} (New York, 1967).

\bibitem{5}N. W. McLachlan {\it Theory and Application of Mathieu Functions} (Oxford Press, 1951).

\bibitem{6}J. A. Stratton {\it Electromagnetic Theory} (Mc-Graw Hill New York, 1941).

\bibitem{7}J. A. Stratton and P. M. Morse {\it Elliptic Cylinder and Spheroidal Wave Functions Including Tables of Separation Constants and Coefficients} (John Wiley \& Sons, New York, 1941).

\bibitem{8}E. T. Whittaker and G. N. Watson {\it A Course of Modern Analysis} (Cambridge University Press, Cambridge, 1950).

\bibitem{9}E. T. Kirkpatrick, ``Tables of values of the modified Mathieu functions,'' Mathematics of Computation {\bf 14} 118--129 (1960).

\bibitem{10}J. C. Gutierrez-Vega, \href{http://homepages.mty.itesm.mx/jgutierr/}
{Formal analysis of the propagation of invariant optical fields in elliptic coordinates}, Ph. D. Thesis, INAOE, Mexico, 2000.

\bibitem{11}J. J. Stamnes and B. Spjelkavik ``New method for computing eigenfunctions (Mathieu functions) for scattering by elliptical cylinders,'' Pure Appl. Opt. {\bf 4} 251--62 (1995).

\bibitem{12}J. J. Stamnes ``Exact two-dimensional scattering by perfectly reflecting elliptical cylinders, strips and slits,'' Pure Appl. Opt. {\bf 4} 841--55 (1995).

\bibitem{13}E. Cojocaru, Matlab free available computer code \href{http://www.mathworks.com/matlabcentral/fileexchange}
{Mathieu Functions Toolbox v. 1.0}; also free available by request at ecojocaru@theory.nipne.ro or cojocaru.e@gmail.com

\bibitem{14}E. Cojocaru, ``Mathieu functions approach to bidimensional scattering by dielectric elliptical cylinders,'' arXiv:0808.2123v1. 

\bibitem{15}E. Cojocaru, ``Elliptical cylindrical invisibility cloak, a semianalytical approach using Mathieu functions,'' arXiv:0808.1498v1.
 
\end{thebibliography}
\end{document}